\documentclass{article} 
\usepackage{iclr2027_conference,times}

\usepackage{hyperref}
\usepackage{url}
\usepackage{graphicx}
\usepackage{booktabs}
\usepackage{amsmath,amssymb}

\title{Debate, Not Voting: Agentic Arbitration\\over Weak Optical Chemical Structure Recognition Models}

\author{
Paul Chauvin \\
Sanofi Digital R\&D, Paris, France \\
\And
Sven Jager \\
Sanofi Digital R\&D, Frankfurt, Germany
}
\iclrfinalcopy 
\begin{document}

\maketitle

\begin{abstract}
Optical chemical structure recognition (OCSR) turns a picture of a molecule into a
machine-readable string, and it stands between decades of printed chemistry and any
computational use of it. On real documents the individual models remain unreliable. Over a
1000-molecule sample of MolParser, three recent readers score $46.0\%$, $43.4\%$ and
$42.4\%$ exact match, yet at least one of them is right on $60.8\%$ of molecules. That gap
is an ensembling problem, but not one voting can solve, because the readers seldom agree
and seldom err together. We instead place the readers inside a multi-agent scaffold. Four
language-model agents with cheminformatics tools argue about a candidate structure, write
their findings to an append-only knowledge base, and are held to render-and-compare gates
that admit a correction only when it is demonstrably better. The debate reaches $61.9\%$,
beating the best single reader by $15.9$ points and plurality voting by $9.2$ points
($p<10^{-9}$ for both, robust to correction for all $27$ tests we report), and is
statistically indistinguishable from the any-reader-right ceiling. That aggregate hides two
opposing effects. On the $661$ molecules carrying no stereochemistry the debate is $3.9$
points above the ceiling ($p=0.004$), constructing structures no reader proposed, while on
the $339$ stereochemical ones it falls below. We then swap the arbitrating model. A stronger
arbitrator adds $2.2$ points (nominal $p=0.039$; neither this nor the $+3.9$ survives
Bonferroni correction over all $27$ tests, so we treat both as suggestive), and its
mechanism is visible: it opens a correction on $423$ of $1000$ molecules against $120$, and
an interaction test indicates that it depends less on the individual readers (nominal
$p=0.018$, not correction-robust). Our conclusion is that the useful unit of progress here
is the arbitration policy rather than the reader, and that OCSR accuracy is better read
against the union oracle than against the best single model.
\end{abstract}

\section{Introduction}

Most of what chemistry knows is drawn, not tabulated. Patents, journal articles and internal
reports depict molecules as line drawings. To use any of it computationally you first need
the structure as a string, usually SMILES \citep{smiles1988}, and that string has to be
recovered from the pixels. The task is called optical chemical structure recognition
(OCSR), and deep learning has improved it a great deal \citep{molscribe2023, decimer2021,
molgrapher2023, molsight2026}. The improvement is uneven, though. On clean synthetic depictions the current
readers are strong. On the drawing styles, wildcards and stereochemical annotations of real
documents they are not.

What limits them is less their average accuracy than where their errors fall. The readers
are weak in different places. On the sample studied here the best of three is right on
$46.0\%$ of molecules, while the union of all three, meaning the molecules where at least
one of them is right, covers $60.8\%$. A reserve of $14.8$ points is therefore already
sitting in the ensemble, unselected. Majority voting recovers about half of it. The readers
disagree too often for a vote to have a majority to count, and where they do agree they tend
to agree on the same mistake.

We ask how much of that reserve a language-model agent system can recover, and which part of
the system does the recovering. Our scaffold treats the readers as frozen, untrusted
witnesses. Four agents exchange evidence through an append-only knowledge base: a structural
analyst that looks at the image, a validator that re-derives properties with RDKit
\citep{rdkit}, a chemist that may propose a correction, and a supervisor that decides.
Render-and-compare gates then refuse any correction that does not survive being drawn again
and checked. The result is closer to adversarial review than to an ensemble average, in the
spirit of debate-based reasoning \citep{du2023debate, irving2018debate} and tool-using
agents \citep{yao2023react}.

Because the scaffold is fixed, we can also vary the arbitrating model alone, which
turns out to change not just accuracy but how much the system depends on any
individual reader.

\paragraph{Contributions.}
\begin{itemize}
\item A debate-based arbitration scaffold over frozen OCSR readers, using an
append-only knowledge base, RDKit tool calls and render-and-compare admission gates
(Section~\ref{sec:method}).
\item A paired evaluation on $1000$ molecules showing that arbitration dominates
every realizable selection rule, and is statistically indistinguishable from the
union-oracle ceiling, with exact McNemar tests throughout (Section~\ref{sec:main}).
\item A decomposition showing the aggregate is a weighted average of two opposite effects,
above the ceiling on non-stereochemical molecules and below it on stereochemical ones
(Section~\ref{sec:where}).
\item An arbitrator-capacity study, with an explicit interaction test showing that a
stronger arbitrator makes the system significantly less dependent on any individual
reader, and identifying its intervention rate as the mechanism
(Sections~\ref{sec:arbitrator}--\ref{sec:mechanism}).
\end{itemize}

\section{Related work}

\paragraph{OCSR.} Rule-based systems such as OSRA \citep{osra2009} gave way to neural
image-to-sequence and image-to-graph models. Img2Mol \citep{img2mol2021} regresses
into a learned latent space; DECIMER \citep{decimer2020, decimer2021} treats the task
as image captioning with a transformer decoder; MolScribe \citep{molscribe2023}
predicts an explicit atom/bond graph, which makes its stereochemistry and its
confidence more interpretable; MolGrapher \citep{molgrapher2023} builds a graph over
detected atoms. More recent work targets exactly the conditions that break these
models in practice: MolSight \citep{molsight2026} adds SMILES pretraining,
multi-granularity supervision and a reinforcement-learning stage aimed at chemical
validity. All of these produce a
single answer per image. Our work is orthogonal: we hold such readers fixed and study
what a reasoning layer above them can recover.

\paragraph{Ensembling and debate.} Classical ensembling assumes errors independent enough to
average out \citep{dietterich2000ensemble}. Here the output is a structured string, so
averaging is undefined and voting is the only obvious aggregation. Within language models,
self-consistency \citep{wang2023selfconsistency} votes over sampled reasoning paths, and
multi-agent debate \citep{du2023debate} lets model instances critique each other, which
improves factuality where a single chain of thought \citep{wei2022cot} fails. Related work
uses models to evaluate other models' output \citep{zheng2023judge} and self-critique to
retry tasks \citep{shinn2023reflexion}. Our setting differs in one respect that matters: the
debaters are not the generators. The candidates come from frozen specialist vision models,
and the agents adjudicate between them against an image, with verification tools to hand.
That makes the union oracle the ceiling on the selection half of the problem, and, as we
show, a ceiling that can be passed where the arbitrator also synthesizes.

\section{Method}
\label{sec:method}

\begin{figure}[t]
\begin{center}
\includegraphics[width=\linewidth]{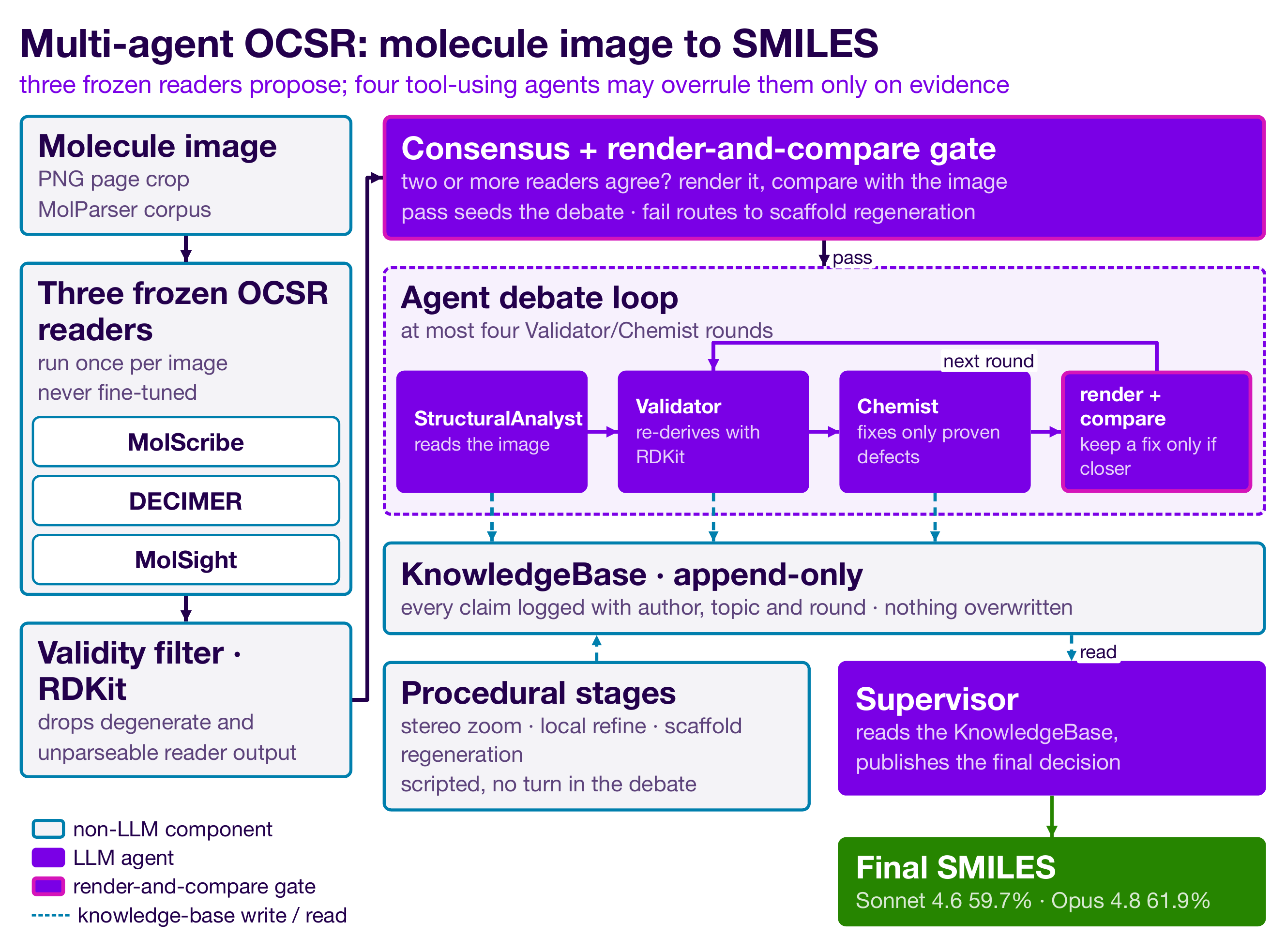}
\end{center}
\caption{The arbitration scaffold. Three frozen readers propose candidate SMILES;
four tool-using agents adjudicate against the image through an append-only knowledge
base; procedural stages contribute targeted evidence; gates admit a correction only
if it survives re-rendering. Only the arbitrating model changes between our two
experimental conditions. The diagram draws the write edges of the three debating agents
plus the Supervisor's read edge, and collapses the three procedural stages into one box.}
\label{fig:arch}
\end{figure}

\subsection{Readers} Figure~\ref{fig:arch} gives the whole scaffold. Three OCSR models run
once per image and are never fine-tuned or otherwise adapted: MolScribe
\citep{molscribe2023}, DECIMER \citep{decimer2021} and MolSight \citep{molsight2026}. Their
outputs are the only structural hypotheses that enter the system. The agents may propose
others, but always with the readers' proposals on the table. A cheap up-front filter drops
degenerate outputs, such as empty strings and unparseable fragments, so that the debate is
not seeded with noise.

\subsection{Agents, tools and the knowledge base}
Four agents are instantiated as tool-using sessions over a shared, append-only
knowledge base. Every message an agent emits is published to the base with an author,
a topic and a round index, and no entry is ever deleted or rewritten; the base is the
full audit trail of a decision, and it is what we mine in
Section~\ref{sec:mechanism}.

\begin{itemize}
\item \textbf{StructuralAnalyst} looks at the image itself and describes what it sees:
ring systems, atom counts, functional groups, stereo annotations. It is not shown a
preferred answer first.
\item \textbf{Validator} re-derives those properties from each candidate SMILES with
RDKit and reports where a candidate contradicts the image description.
\item \textbf{Chemist} may propose a corrected structure, but only against a
specific, stated defect.
\item \textbf{Supervisor} issues the final decision, which is either to keep the
incumbent candidate or to adopt a proposed replacement.
\end{itemize}

Rather than assert chemistry from memory, the agents call deterministic cheminformatics
tools: SMILES validation, atom and ring counts, functional-group and stereocentre
extraction, pairwise property comparison, and rendering a SMILES back to an image. That last
tool is what makes the gates below possible.

Three further roles are procedural rather than conversational. Each has fixed control flow,
issues its own narrowly scoped vision queries, and publishes evidence to the knowledge base,
but takes no turn in the debate. A StereoSpecialist resolves stereocentres by zooming into
the relevant region of the image, a LocalRefiner zooms into the region where the readers
disagree, and a Generator regenerates a structure from a maximum-common-substructure
scaffold when no candidate is credible. We spell this split out because it is easy to
overstate how much of a system like this is really agentic. Four roles argue; three follow a
script.

\subsection{The loop and its gates} The Validator and Chemist alternate for up to four
rounds, after which the Supervisor decides. Debate by itself is no safeguard, since a fluent
argument for a wrong structure is still wrong. Admission is therefore governed by gates that
can veto a change however well it was argued:

\begin{itemize}
\item \textbf{Render-and-compare on consensus.} When the readers agree, the agreed
structure is rendered and checked against the original image before being accepted.
\item \textbf{Render-and-compare on corrections.} A Chemist fix must be re-rendered
and compared to the image; a fix that does not visibly match is discarded.
\item \textbf{Protected unanimity.} The Chemist may not overturn a structure that all
readers agreed on. This gate exists because, in development, overturning unanimous
agreement degraded results without ever rescuing a wrong answer.
\item \textbf{Markush pass-through.} A reader's R-group wildcard short-circuits the
debate instead of being argued away into a concrete structure. This gate is nearly inert
here, since our readers emit a wildcard on only $17$ of the $163$ Markush molecules.
\end{itemize}

Each of these components is individually switchable, which is how we ablate them.

\paragraph{The single free variable.} All agent and procedural roles are served by one
vision-language model, named by a single configuration constant. Our second experiment
changes that constant and nothing else in the scaffold, so reader outputs, prompts,
gates, round limits and scoring are identical across conditions; we name the two
settings by their arbitrator, \emph{Sonnet 4.6} (\texttt{eu.anthropic.claude-sonnet-4-6})
and \emph{Opus 4.8} (\texttt{eu.anthropic.claude-opus-4-8}), both served as hosted
Amazon Bedrock inference profiles and run in September 2026. One difference we did not
choose follows from that swap, because the second model rejects the sampling-temperature
parameter; we quantify it in Section~\ref{sec:limits}.

\section{Experimental setup}

\paragraph{Data.} We sample $1000$ molecules uniformly at random from the MolParser
corpus \citep{molparser2025}, which is drawn from real documents and therefore retains
the properties that make OCSR hard: varied drawing conventions, salts, explicit
stereochemistry, and Markush R-group wildcards. $163$ of the $1000$ ground-truth
structures contain a wildcard and $339$ carry stereochemistry. One ground-truth string
does not parse under RDKit; we retain it and count it wrong for every system rather
than silently dropping it.

\paragraph{Metric.} Exact match after canonicalisation. A prediction counts as correct if
RDKit parses it and its canonical SMILES equals the canonical ground truth, stereochemistry
included. We chose a strict criterion on purpose, because it is the one under which a
recovered structure can be trusted as a database record.

\paragraph{Baselines.} Beyond the individual readers we report two selection rules and one
bound. \emph{Plurality vote} takes the most frequent canonical answer, breaking ties by a
fixed reader priority (MolScribe $>$ DECIMER $>$ MolSight, i.e.\ decreasing standalone
accuracy). With three readers that all disagree it falls back on the priority reader, which
is why it outscores that reader overall, $52.7\%$ against $46.0\%$. It does not beat it
molecule by molecule: it loses the five cases where the two lower-priority readers agree on
the same wrong answer. It is still the stronger of the two voting variants. \emph{Strict
majority} requires at least two readers to agree and otherwise abstains, counting as wrong.
The \emph{union oracle} marks a molecule correct if any reader is correct. No system can
implement the oracle, since picking the right reader requires knowing the answer, but it is
the exact ceiling on any procedure that only ever selects among reader outputs, and so the
right reference for an arbitration layer.

\paragraph{Statistics.} All systems are evaluated on the same molecules, and the
readers are identical across both arbitrated runs, so every comparison is paired. We
report Wilson score intervals \citep{wilson1927} for accuracies and, for differences, a
two-sided exact McNemar test \citep{mcnemar1947} on the discordant pairs together with
a paired bootstrap interval over $20{,}000$ resamples. We verified that base
predictions are identical across the two runs (zero mismatches over $1000$ molecules)
and recomputed every correctness flag from the stored SMILES rather than trusting the
pipeline's own bookkeeping (zero disagreements).

\paragraph{Ablations.} Leave-one-reader-out requires re-running the whole pipeline per
configuration, so it is done on a fixed subset of $300$ molecules, with the full
pipeline re-measured on that same subset as the reference.

\section{Results}

\begin{figure}[t]
\begin{center}
\includegraphics[width=\linewidth]{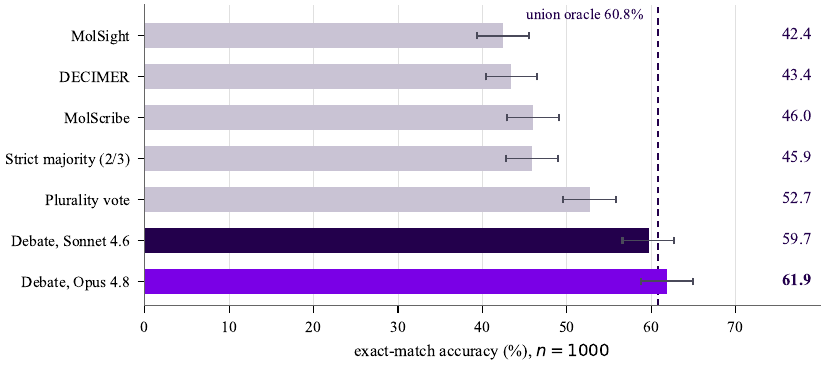}
\end{center}
\caption{Exact-match accuracy with 95\% Wilson intervals, $n=1000$. Both arbitrated
systems clear every selection rule by a wide margin and sit at the union-oracle
ceiling (dashed).}
\label{fig:results}
\end{figure}

\begin{table}[t]
\caption{Exact-match accuracy on the same 1000 MolParser molecules. The three readers
are identical across both arbitrated runs, so every row is measured on the same inputs
and all comparisons are paired. Intervals are 95\% Wilson score intervals.}
\label{tab:main}
\begin{center}
\small
\begin{tabular}{llcc}
\multicolumn{1}{c}{\bf System} & \multicolumn{1}{c}{\bf Type}
& \multicolumn{1}{c}{\bf Correct} & \multicolumn{1}{c}{\bf Accuracy (95\% CI)}
\\ \hline \\
MolSight & single reader & 424 & 42.4\% (39.4--45.5) \\
DECIMER & single reader & 434 & 43.4\% (40.4--46.5) \\
MolScribe & single reader & 460 & 46.0\% (42.9--49.1) \\
Strict majority (2/3) & selection & 459 & 45.9\% (42.8--49.0) \\
Plurality vote & selection & 527 & 52.7\% (49.6--55.8) \\
Debate, Sonnet 4.6 & arbitration & 597 & 59.7\% (56.6--62.7) \\
\bf Debate, Opus 4.8 & arbitration & 619 & 61.9\% (58.8--64.9) \\
Union oracle (ceiling) & upper bound & 608 & 60.8\% (57.7--63.8) \\
\end{tabular}
\end{center}
\end{table}

\begin{table}[t]
\caption{Paired comparisons. $b$ is the number of molecules the first system gets right
and the second wrong, $c$ the reverse; $p$ is from a two-sided exact McNemar test on the
$b+c$ discordant pairs. Differences are in percentage points with a paired bootstrap
95\% interval (20{,}000 resamples).}
\label{tab:paired}
\begin{center}
\small
\begin{tabular}{llcccc}
\multicolumn{1}{c}{\bf System} & \multicolumn{1}{c}{\bf Reference}
& \multicolumn{1}{c}{\bf $\Delta$ (pp)} & \multicolumn{1}{c}{\bf 95\% CI}
& \multicolumn{1}{c}{\bf $b$\,/\,$c$} & \multicolumn{1}{c}{\bf $p$}
\\ \hline \\
Debate, Opus 4.8 & Debate, Sonnet 4.6 & +2.2 & [+0.2, +4.2] & 63\,/\,41 & 0.039 \\
Debate, Opus 4.8 & Union oracle (ceiling) & +1.1 & [-1.2, +3.4] & 76\,/\,65 & 0.400 \\
Debate, Sonnet 4.6 & Union oracle (ceiling) & -1.1 & [-3.1, +0.9] & 46\,/\,57 & 0.324 \\
Debate, Opus 4.8 & MolScribe & +15.9 & [+13.2, +18.7] & 189\,/\,30 & $<$0.0001 \\
Debate, Sonnet 4.6 & MolScribe & +13.7 & [+11.1, +16.3] & 166\,/\,29 & $<$0.0001 \\
Debate, Opus 4.8 & Plurality vote & +9.2 & [+6.8, +11.7] & 126\,/\,34 & $<$0.0001 \\
Debate, Sonnet 4.6 & Plurality vote & +7.0 & [+4.9, +9.2] & 97\,/\,27 & $<$0.0001 \\
Debate, Opus 4.8 & Strict majority (2/3) & +16.0 & [+13.5, +18.6] & 178\,/\,18 & $<$0.0001 \\
\end{tabular}
\end{center}
\end{table}

\begin{table}[t]
\caption{Accuracy by molecule class. Markush molecules carry an R-group wildcard that
the readers emit only rarely (on $17$ of the $163$), so the whole stack is near zero there.
On molecules that carry stereochemistry both arbitrators fall below the union oracle; on
the rest Opus~4.8 rises above it.}
\label{tab:subgroups}
\begin{center}
\small
\begin{tabular}{lccccc}
\multicolumn{1}{c}{\bf Subset} & \multicolumn{1}{c}{\bf $n$}
& \multicolumn{1}{c}{\bf MolScribe} & \multicolumn{1}{c}{\bf Plurality}
& \multicolumn{1}{c}{\bf Oracle} & \multicolumn{1}{c}{\bf Sonnet / Opus}
\\ \hline \\
Contains a Markush wildcard & 163 & 3.7\% & 3.1\% & 3.7\% & 0.0\% / 1.2\% \\
No Markush wildcard & 837 & 54.2\% & 62.4\% & 71.9\% & 71.3\% / 73.7\% \\
Carries stereochemistry & 339 & 35.1\% & 40.1\% & 49.0\% & 44.8\% / 44.5\% \\
No stereochemistry & 661 & 51.6\% & 59.2\% & 66.9\% & 67.3\% / 70.8\% \\
\end{tabular}
\end{center}
\end{table}

\begin{table}[t]
\caption{Paired tests against the union oracle \emph{within} each subset. Aggregated over
all 1000 molecules the debate merely matches the ceiling, but that average hides two
opposing effects: on the 661 molecules with no stereochemistry Opus~4.8 is significantly
\emph{above} it, and on the 339 that carry stereochemistry both arbitrators fall below.}
\label{tab:subgrouptests}
\begin{center}
\small
\begin{tabular}{llcccc}
\multicolumn{1}{c}{\bf Subset} & \multicolumn{1}{c}{\bf Comparison}
& \multicolumn{1}{c}{\bf $\Delta$ (pp)} & \multicolumn{1}{c}{\bf 95\% CI}
& \multicolumn{1}{c}{\bf $b$\,/\,$c$} & \multicolumn{1}{c}{\bf $p$}
\\ \hline \\
No Markush ($n=837$) & Opus 4.8 vs.\ oracle & +1.8 & [-1.0, +4.5] & 76\,/\,61 & 0.232 \\
No Markush ($n=837$) & Sonnet 4.6 vs.\ oracle & -0.6 & [-2.9, +1.8] & 46\,/\,51 & 0.685 \\
No Markush ($n=837$) & Opus 4.8 vs.\ Sonnet 4.6 & +2.4 & [+0.0, +4.8] & 61\,/\,41 & 0.059 \\
Stereo ($n=339$) & Opus 4.8 vs.\ oracle & -4.4 & [-9.1, +0.3] & 25\,/\,40 & 0.082 \\
Stereo ($n=339$) & Sonnet 4.6 vs.\ oracle & -4.1 & [-8.0, -0.3] & 17\,/\,31 & 0.059 \\
Stereo ($n=339$) & Opus 4.8 vs.\ Sonnet 4.6 & -0.3 & [-4.7, +4.1] & 27\,/\,28 & 1.000 \\
No stereo ($n=661$) & Opus 4.8 vs.\ oracle & +3.9 & [+1.4, +6.5] & 51\,/\,25 & 0.004 \\
No stereo ($n=661$) & Sonnet 4.6 vs.\ oracle & +0.5 & [-1.8, +2.7] & 29\,/\,26 & 0.788 \\
No stereo ($n=661$) & Opus 4.8 vs.\ Sonnet 4.6 & +3.5 & [+1.5, +5.6] & 36\,/\,13 & 0.001 \\
\end{tabular}
\end{center}
\end{table}

\subsection{Arbitration beats selection and matches the ceiling}
\label{sec:main}

Table~\ref{tab:main} and Figure~\ref{fig:results} give the main picture. The three readers
cluster between $42.4\%$ and $46.0\%$, and no pair of their intervals is far apart. These
are three comparably capable models that disagree. Plurality voting reaches $52.7\%$, which
recovers about $6.7$ of the $14.8$ points between the best reader and the oracle. Requiring
a genuine two-of-three majority does worse than the best single reader, $45.9\%$ against
$46.0\%$. Agreement is scarce in this regime, and that comparison shows it most plainly.

Both arbitrated systems sit well above all of that. Opus~4.8 reaches $61.9\%$ and Sonnet~4.6
$59.7\%$. Against the best single reader those are gains of $15.9$ and $13.7$ points, and
against plurality voting $9.2$ and $7.0$ points, all at $p<10^{-9}$
(Table~\ref{tab:paired}). The discordance counts confirm this is not a wash. Opus~4.8
rescues $189$ molecules that MolScribe gets wrong, and loses $30$ that it gets right.

The comparison with the union oracle needs more care, because a favourable number here is
easy to over-read. Opus~4.8's $61.9\%$ is nominally above the oracle's $60.8\%$, but the
paired test gives $+1.1$ points with a $95\%$ interval of $[-1.2, +3.4]$ and $p=0.40$. What
we can say is that the debate is \emph{statistically indistinguishable} from the ceiling on
any procedure that selects among reader outputs. We have not shown equivalence, and the
interval still allows the debate to sit $1.2$ points below it. Since no system can reach the
ceiling in practice, arriving in its neighbourhood is the operationally useful result. It is
not the same thing as beating it, and at $n=1000$ the stronger claim is out of reach in
aggregate.

\subsection{Where arbitration wins, and what it costs}
\label{sec:where}

That aggregate average conceals two effects that pull in opposite directions
(Tables~\ref{tab:subgroups} and~\ref{tab:subgrouptests}).

On the $661$ molecules with no stereochemistry, Opus~4.8 reaches $70.8\%$ against an oracle
of $66.9\%$: $+3.9$ points, $95\%$ interval $[+1.4, +6.5]$, $p=0.004$. Here the system does
more than pick a reader. It assembles correct structures that none of its three readers
produced. Within this subset it answers $51$ molecules correctly where every reader was wrong and
loses $25$ that some reader had right, a net $+26$; on the stereochemical molecules the same
counts run the other way, $25$ against $40$, a net $-15$. The full-sample net of $+11$ is
what survives that cancellation.
Two caveats belong with this result. The stereochemistry split was chosen after we saw the
aggregate, not pre-registered, and it is the largest positive effect among nine subgroup
tests. It is the finding we would most want replicated, not one to build on unreplicated.

On the $339$ molecules that carry stereochemistry, both arbitrators sit below the oracle, at
$44.5\%$ and $44.8\%$ against $49.0\%$. The residual error profile shows where it goes.
Relative to Sonnet~4.6, Opus~4.8 removes $27$ skeletal errors and $5$ invalid strings, but
adds $16$ stereochemical ones (Table~\ref{tab:taxonomy}). Arguing about constitution and
arguing about chirality are not the same skill, and a debate that improves the first can
damage the second. Neither stereo deficit reaches significance under the exact McNemar test
($p=0.08$ and $p=0.06$), though the bootstrap interval for Sonnet~4.6 does exclude zero.
Where the two disagree we defer to the test, and treat the direction as consistent rather
than established.

Markush structures are a hard floor for the whole stack. On the $163$ molecules whose ground
truth carries an R-group wildcard, the readers' union is right $3.7\%$ of the time and the
arbitrated systems $0.0\%$ and $1.2\%$. The readers almost never emit a wildcard at all, so
there is usually nothing for arbitration to select. These molecules are $16.3\%$ of the
sample, and they pull every aggregate number here down by $8$ to $12$ points
(Appendix~\ref{app:markush}).

The residual errors are near-misses, not nonsense. The median Tanimoto similarity between a
wrong final answer and the ground truth is $0.72$ under Sonnet~4.6 and $0.76$ under
Opus~4.8, and $999$ of $1000$ final answers under Opus~4.8 are valid chemical structures.

\subsection{Swapping the arbitrator}
\label{sec:arbitrator}

We hold the scaffold, prompts, gates and reader outputs fixed, so what differs between the
two conditions is the arbitrating \emph{configuration}: the model, and the sampling
temperature that model forced on us (Section~\ref{sec:limits}). Opus~4.8 gains $2.2$ points
over Sonnet~4.6, with a $95\%$ interval of $[+0.2, +4.2]$, $p=0.039$ and discordance
$63/41$. The effect is real but modest, and it lands where the subgroup analysis says it
should. On non-stereochemical molecules the gap is $+3.5$ points ($p=0.001$); on
stereochemical ones it is $-0.3$ points ($p=1.0$). A stronger arbitrator buys constitutional
reasoning. For stereochemistry it buys nothing at all.

\subsection{What the arbitrator's capacity actually changes}
\label{sec:mechanism}

\begin{figure}[t]
\begin{center}
\includegraphics[width=\linewidth]{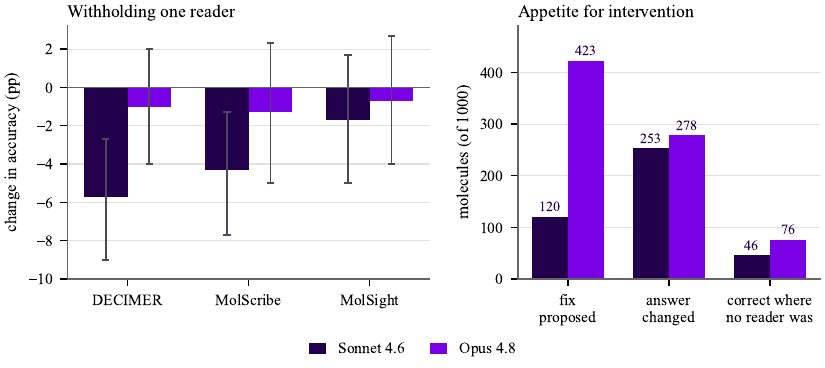}
\end{center}
\caption{Left: change in accuracy when one reader is withheld and the pipeline re-run
($n=300$, 95\% paired bootstrap intervals). Under Sonnet~4.6 two readers are
load-bearing; under Opus~4.8 every interval crosses zero. Right: how often each
arbitrator opens a correction, changes the answer, and is right where no reader was.}
\label{fig:ablation}
\end{figure}

Because every claim is logged, we can ask what the stronger arbitrator does differently, not
just what it scores. It intervenes far more (Table~\ref{tab:behaviour},
Figure~\ref{fig:ablation} right). Opus~4.8 opens a correction on $423$ of $1000$ molecules
where Sonnet~4.6 opens one on $120$, a factor of $3.5$. (A molecule counts as having a
correction opened if \emph{any} agent published a proposed fix. By author the increase is
$78 \to 143$ for the Chemist and $78 \to 377$ for the StructuralAnalyst, so it is mainly the
agent that reads the image that raises objections, not the one designated to repair the
structure.) Yet the final answer differs from the seeded candidate on only $278$ molecules
against $253$. Most of the extra proposals are talked down by the system's own gates and
validators, which is the gates doing their job. An arbitrator with a large appetite for
correction is only as safe as whatever can veto it.

The net effect decomposes cleanly. Opus~4.8 is correct where no reader was on $76$ molecules
against $46$, a gain of $30$, and loses $65$ winnable molecules against $57$, a loss of $8$.
The difference of $+22$ molecules is exactly the $2.2$-point aggregate gap. Intervening more
is not free. The per-molecule counts are much closer than the net figures suggest, $76$
rescues against $65$ losses, and the policy pays here only because its \emph{net} gain is
about four times its net loss.

The readers' role changes too. Withholding one reader and re-running the full pipeline
(Table~\ref{tab:ablation}, Figure~\ref{fig:ablation} left) costs $5.7$ and $4.3$ points
under Sonnet~4.6 for DECIMER and MolScribe, both significant, while MolSight is worth $1.7$
points and is not. Under Opus~4.8 no reader is load-bearing. The three deficits are $1.0$,
$1.3$ and $0.7$ points, every interval crosses zero, and $p \ge 0.58$ throughout.

Landing on opposite sides of a significance threshold is not by itself evidence of a
difference, so we test the interaction directly. On the same $300$ molecules we take, for
each withheld reader, the paired difference-in-differences between the two arbitrators'
deficits, and assess it by permuting the arbitrator label within each molecule ($20{,}000$
permutations, fixed seed). Pooled over the three readers, withholding a reader costs
Sonnet~4.6 $2.9$ points more than it costs Opus~4.8 ($p=0.018$ over $900$ reader-molecule
pairs). Per reader the contrast is $4.7$ points for DECIMER ($p=0.027$) and not significant
for the other two. So the reduction in reader dependence is a measured effect, not merely an
absence of evidence, although at this sample size we establish it only in aggregate. A more
capable arbitrator substitutes its own reading for whichever reader is missing. That is
convenient operationally, since the reader set can be trimmed for cost, but it also means
reader-level progress buys less as the arbitration layer improves.

\section{Limitations}
\label{sec:limits}

\paragraph{The two conditions differ by more than the model.} Opus~4.8 rejects the
sampling-temperature parameter through the API we used. Our scaffold requests temperature
$0$ on every call, so Sonnet~4.6 runs near-deterministically, whereas under Opus~4.8 the
parameter was dropped and every call ran at the provider default. This is a global
difference, not a per-component one. We therefore cannot claim the model is the only free
variable, and we do not know the sign of the effect of sampling one arm stochastically and
the other near-greedily. One consequence is bounded. The scaffold-regeneration path takes a
three-sample self-consistency vote that degenerates to a single sample without a
temperature, but that path fires on only $7$ of $1000$ molecules in either condition, so it
can account for at most $0.7$ points. The wider exposure, that every other call was sampled
differently, our data cannot bound. A properly matched replication would pin the temperature
explicitly in both arms.

\paragraph{Multiplicity.} We report $27$ paired tests: eight system comparisons, nine
subgroup tests, six leave-one-out tests and four interaction tests. A Bonferroni threshold
over the whole family ($0.05/27 \approx 0.0019$) leaves seven. Those are the five
comparisons against the single readers and the voting rules, all below $10^{-9}$, plus the
non-stereochemical arbitrator gap ($p=0.001$) and the DECIMER ablation under Sonnet~4.6
($p<0.001$). Three claims are more fragile, and we mark them as such wherever they appear:
the $+2.2$-point arbitrator gap ($p=0.039$), the pooled interaction test ($p=0.018$), and
the $+3.9$ points above the oracle on non-stereochemical molecules ($p=0.004$, which
survives within the nine-test subgroup family but not across all $27$). Bonferroni is
conservative for tests this correlated, so these thresholds are pessimistic rather than
sharp.

\paragraph{Scope of the evidence.} All results come from one $1000$-molecule sample of one
corpus, one run per condition. Our intervals resample molecules, so they capture sampling
uncertainty but not the agents' own stochasticity. For the arbitrator comparison, where one
arm ran at a non-zero temperature, some of the $63/41$ discordance is probably run-to-run
noise that our tests attribute to the model, and we would want a condition repeated on the
$300$-molecule subset before the $+2.2$-point gap is relied upon. We also do not know
whether the readers' training data overlaps this corpus. Three baselines a reader will want
are missing: a zero-reader arm, an ablation of the scaffold's own components, and a fourth
reader able to emit R-group wildcards (Appendix~\ref{app:omitted}). Finally, our three
readers are comparably capable, whereas a configuration containing a dominant one would
behave differently.

\paragraph{Cost, metric and silent error.} The readers still run, so arbitration adds
roughly thirteen knowledge-base entries per molecule, and the model calls behind them, on
top of three forward passes. That is defensible for a one-off corpus extraction and not for interactive
use. Exact match under-credits partially correct structures a chemist could still use, and
over-penalises Markush cases where the wildcard convention is itself ambiguous. The union
oracle bounds selection rather than synthesis, which is why our non-stereochemical result
can exceed it, and why it should not be read as an upper bound on the task. One limitation
matters more than any of these. $999$ of $1000$ outputs are valid parseable structures while
$38.1\%$ are wrong, so essentially every error this system makes is a silent one, and it
offers no abstention.

\section{Conclusion}

Three OCSR readers that individually answer fewer than half of a real-document sample
collectively contain the right answer $60.8\%$ of the time, and no voting rule gets close to
extracting it. A debate among tool-using agents, constrained by render-and-compare gates,
reaches $61.9\%$, which is statistically indistinguishable from that ceiling and $15.9$
points above the best reader. Splitting by stereochemistry shows the system is doing more
than selecting. Without stereochemistry it rises above the ceiling and constructs structures
none of its readers proposed; on stereochemical molecules it damages information a reader
had right. A stronger arbitrator adds $2.2$ points and leans less on any individual reader,
while intervening $3.5\times$ more often, though both effects are nominal rather than
correction-robust. For several comparably weak readers with none dominant, the arbitration
policy rather than the reader is where the remaining accuracy lives.

\subsection*{AI use statement}

In this work, generative AI is the object of study: the arbitration layer evaluated here
consists of hosted large language models (Anthropic Claude Sonnet~4.6 and Opus~4.8),
and the OCSR readers are separately published neural models. All such use is described
in Section~\ref{sec:method} and in the experimental setup.

Separately, generative AI tools were used as coding assistance and for editorial
polishing of the text. All such output was reviewed. The scaffold, the experimental
design and the claims are the authors' own, and generative AI was not used to generate
or alter experimental data. Every quantity reported here is computed by analysis code
directly from the stored per-molecule records rather than transcribed, and the tables
and figures are generated from that output; the statistics were reproduced by a second,
independent implementation, and all cited references were checked against their primary
sources. The authors take responsibility for the final content of this work.

\subsection*{Ethics statement}

This work involves no human subjects and no personally identifying data. The molecular
depictions are drawn from a publicly released OCSR corpus. The intended application,
recovering machine-readable structures from documents, is a data-extraction task. Its main
foreseeable harm is silent error. An incorrectly recovered structure entering a chemical
database is worse than no record at all, because downstream users cannot see that it is
wrong. That is the reason for the strict exact-match metric and for reporting near-miss
similarity in Section~\ref{sec:where}. We would discourage deploying such a system without
either human review or a calibrated abstention mechanism, which the present system does not
provide. Chemical structure recognition is dual-use in the general sense that any chemistry
tooling is, but this work introduces no capability specific to harmful chemistry.

\subsection*{Reproducibility statement}

Section~\ref{sec:method} specifies the agents, tools, gates, round limits and the single
configuration constant that differs between our two conditions. The experimental setup gives
the evaluation protocol, the metric, the statistical procedures, and the baseline
definitions, including the exact tie-breaking rule for plurality voting, which materially
affects that baseline. Appendix~\ref{app:taxonomy} gives the mutually exclusive rule used to
assign every residual error to a class, so the taxonomy can be reproduced exactly.
Appendix~\ref{app:integrity} records the integrity checks. Per-molecule records for both
conditions, including the full knowledge-base transcript for every molecule, together with
the analysis scripts that compute every number in this paper, are available as supplementary
material.

\bibliography{refs}

@inproceedings{molparser2025,
  title     = {{MolParser: End-to-end Visual Recognition of Molecule Structures in the Wild}},
  author    = {Fang, Xi and Wang, Jiankun and Cai, Xiaochen and Chen, Shangqian and
               Yang, Shuwen and Tao, Haoyi and Wang, Nan and Yao, Lin and
               Zhang, Linfeng and Ke, Guolin},
  booktitle = {Proceedings of the IEEE/CVF International Conference on Computer Vision (ICCV)},
  year      = {2025},
  note      = {arXiv:2411.11098}
}

@inproceedings{molsight2026,
  title     = {{MolSight: Optical Chemical Structure Recognition with {SMILES} Pretraining,
               Multi-Granularity Learning and Reinforcement Learning}},
  author    = {Zhang, Wenrui and Wang, Xinggang and Feng, Bin and Liu, Wenyu},
  booktitle = {Proceedings of the AAAI Conference on Artificial Intelligence},
  year      = {2026},
  note      = {arXiv:2511.17300}
}

@article{molscribe2023,
  title   = {{MolScribe: Robust Molecular Structure Recognition with Image-to-Graph Generation}},
  author  = {Qian, Yujie and Guo, Jiang and Tu, Zhengkai and Li, Zhening and
             Coley, Connor W. and Barzilay, Regina},
  journal = {Journal of Chemical Information and Modeling},
  volume  = {63},
  number  = {7},
  pages   = {1925--1934},
  year    = {2023}
}

@article{decimer2021,
  title   = {{DECIMER 1.0: deep learning for chemical image recognition using transformers}},
  author  = {Rajan, Kohulan and Zielesny, Achim and Steinbeck, Christoph},
  journal = {Journal of Cheminformatics},
  volume  = {13},
  number  = {1},
  pages   = {61},
  year    = {2021}
}

@article{decimer2020,
  title   = {{DECIMER: towards deep learning for chemical image recognition}},
  author  = {Rajan, Kohulan and Zielesny, Achim and Steinbeck, Christoph},
  journal = {Journal of Cheminformatics},
  volume  = {12},
  number  = {1},
  pages   = {65},
  year    = {2020}
}

@inproceedings{molgrapher2023,
  title     = {{MolGrapher: Graph-based Visual Recognition of Chemical Structures}},
  author    = {Morin, Lucas and Danelljan, Martin and Agea, Maria Isabel and
               Nassar, Ahmed and Weber, Valery and Meijer, Ingmar and
               Staar, Peter and Yu, Fisher},
  booktitle = {Proceedings of the IEEE/CVF International Conference on Computer Vision (ICCV)},
  year      = {2023}
}

@article{img2mol2021,
  title   = {{Img2Mol -- accurate {SMILES} recognition from molecular graphical depictions}},
  author  = {Clevert, Djork-Arn{\'e} and Le, Tuan and Winter, Robin and Montanari, Floriane},
  journal = {Chemical Science},
  volume  = {12},
  number  = {42},
  pages   = {14174--14181},
  year    = {2021}
}

@article{osra2009,
  title   = {{Optical structure recognition software to recover chemical information:
             {OSRA}, an open source solution}},
  author  = {Filippov, Igor V. and Nicklaus, Marc C.},
  journal = {Journal of Chemical Information and Modeling},
  volume  = {49},
  number  = {3},
  pages   = {740--743},
  year    = {2009}
}

@article{smiles1988,
  title   = {{SMILES, a chemical language and information system. 1.
             Introduction to methodology and encoding rules}},
  author  = {Weininger, David},
  journal = {Journal of Chemical Information and Computer Sciences},
  volume  = {28},
  number  = {1},
  pages   = {31--36},
  year    = {1988}
}

@misc{rdkit,
  title        = {{RDKit: Open-source cheminformatics}},
  author       = {Landrum, Greg and others},
  howpublished = {\url{https://www.rdkit.org}},
  year         = {2025},
  note         = {Release 2025.09}
}

@inproceedings{du2023debate,
  title     = {{Improving Factuality and Reasoning in Language Models
               through Multiagent Debate}},
  author    = {Du, Yilun and Li, Shuang and Torralba, Antonio and
               Tenenbaum, Joshua B. and Mordatch, Igor},
  booktitle = {Proceedings of the International Conference on Machine Learning (ICML)},
  year      = {2024}
}

@inproceedings{wang2023selfconsistency,
  title     = {{Self-Consistency Improves Chain of Thought Reasoning in Language Models}},
  author    = {Wang, Xuezhi and Wei, Jason and Schuurmans, Dale and Le, Quoc V. and
               Chi, Ed H. and Narang, Sharan and Chowdhery, Aakanksha and Zhou, Denny},
  booktitle = {International Conference on Learning Representations (ICLR)},
  year      = {2023}
}

@inproceedings{wei2022cot,
  title     = {{Chain-of-Thought Prompting Elicits Reasoning in Large Language Models}},
  author    = {Wei, Jason and Wang, Xuezhi and Schuurmans, Dale and Bosma, Maarten and
               Ichter, Brian and Xia, Fei and Chi, Ed H. and Le, Quoc V. and Zhou, Denny},
  booktitle = {Advances in Neural Information Processing Systems (NeurIPS)},
  year      = {2022}
}

@inproceedings{yao2023react,
  title     = {{ReAct: Synergizing Reasoning and Acting in Language Models}},
  author    = {Yao, Shunyu and Zhao, Jeffrey and Yu, Dian and Du, Nan and
               Shafran, Izhak and Narasimhan, Karthik and Cao, Yuan},
  booktitle = {International Conference on Learning Representations (ICLR)},
  year      = {2023}
}

@inproceedings{shinn2023reflexion,
  title     = {{Reflexion: Language Agents with Verbal Reinforcement Learning}},
  author    = {Shinn, Noah and Cassano, Federico and Gopinath, Ashwin and
               Narasimhan, Karthik and Yao, Shunyu},
  booktitle = {Advances in Neural Information Processing Systems (NeurIPS)},
  year      = {2023}
}

@inproceedings{zheng2023judge,
  title     = {{Judging {LLM-as-a-Judge} with {MT-Bench} and Chatbot Arena}},
  author    = {Zheng, Lianmin and Chiang, Wei-Lin and Sheng, Ying and Zhuang, Siyuan and
               Wu, Zhanghao and Zhuang, Yonghao and Lin, Zi and Li, Zhuohan and
               Li, Dacheng and Xing, Eric P. and Zhang, Hao and
               Gonzalez, Joseph E. and Stoica, Ion},
  booktitle = {Advances in Neural Information Processing Systems (NeurIPS)
               Datasets and Benchmarks Track},
  year      = {2023}
}

@misc{irving2018debate,
  title  = {{AI safety via debate}},
  author = {Irving, Geoffrey and Christiano, Paul and Amodei, Dario},
  year   = {2018},
  note   = {arXiv:1805.00899}
}

@incollection{dietterich2000ensemble,
  title     = {{Ensemble Methods in Machine Learning}},
  author    = {Dietterich, Thomas G.},
  booktitle = {Multiple Classifier Systems},
  series    = {Lecture Notes in Computer Science},
  volume    = {1857},
  pages     = {1--15},
  publisher = {Springer},
  year      = {2000}
}

@article{mcnemar1947,
  title   = {{Note on the sampling error of the difference between correlated
             proportions or percentages}},
  author  = {McNemar, Quinn},
  journal = {Psychometrika},
  volume  = {12},
  number  = {2},
  pages   = {153--157},
  year    = {1947}
}

@article{wilson1927,
  title   = {{Probable Inference, the Law of Succession, and Statistical Inference}},
  author  = {Wilson, Edwin B.},
  journal = {Journal of the American Statistical Association},
  volume  = {22},
  number  = {158},
  pages   = {209--212},
  year    = {1927}
}
\bibliographystyle{iclr2027_conference}

\newpage
\appendix
\section{Appendix}

\subsection{Error taxonomy rule}
\label{app:taxonomy}

Every molecule the system answers incorrectly is assigned to exactly one class, by the
first matching rule in this order. Let $g$ be the ground-truth SMILES and $f$ the final
answer.

\begin{enumerate}
\item \textbf{No valid output} --- $f$ is absent or does not parse under RDKit.
\item \textbf{Markush R-group} --- $g$ and $f$ contain different numbers of \texttt{*}
wildcard atoms.
\item \textbf{Stereochemistry} --- $g$ and $f$ are equal once stereochemistry is
stripped, i.e.\ the constitution is right and only the stereo descriptors differ.
\item \textbf{Salt / counter-ion} --- the multisets of dot-separated fragments differ
while the largest fragment is identical.
\item \textbf{Skeleton} --- anything else; the molecular constitution itself is wrong.
\end{enumerate}

Because rule~2 counts wildcards in \emph{either} direction, the Markush error count need
not equal the number of Markush ground truths. Under Sonnet~4.6 the $164$ errors in
Table~\ref{tab:taxonomy} decompose as $162$ Markush ground truths whose wildcard count the
answer did not match, plus $2$ non-Markush ground truths that received an answer containing
a wildcard; the remaining Markush molecule is counted under rule~1 because no parseable
answer was produced at all.

\subsection{Baselines we did not run}
\label{app:omitted}
Three absences are worth naming explicitly. First, we do not report a zero-reader arm in
which the arbitrating model reads the image alone, so we cannot fully separate ``debate
over weak readers works'' from ``this model can read molecule images''; the finding that
no single reader is load-bearing under Opus~4.8 makes that the obvious next experiment,
and the leave-one-out harness already supports it. Second, we do not ablate the scaffold's own
components, meaning the render-and-compare gates, the RDKit tools and the debate loop. Our
contrast with voting is therefore established against selection rules, not against a
tool-using single agent. Third, our readers emit R-group wildcards only
rarely, on $17$ of the $163$ Markush molecules, and that is our worst failure mode
(Appendix~\ref{app:markush}). A reader that emitted them routinely would raise both the
union oracle and, very likely, the system. We held
one reader set fixed across both arbitrator conditions so that the arbitrator comparison
stayed paired, and that choice is what excludes such a reader here.

\subsection{The Markush floor}
\label{app:markush}
The readers rarely produce a wildcard at all: across the $163$ Markush molecules
MolScribe emits one on $16$ and is right on $6$, MolSight emits one on $3$ and is right on
none, and DECIMER never does. That is why the union oracle reaches only $3.7\%$ here.
The debate then occasionally argues a wildcard away into a concrete structure, so it can
score \emph{below} the readers' union on this subset. Because the floor is near zero,
the stronger a system is on the rest of the sample the more these $163$ molecules cost
it: the depression from the non-Markush subset to the full sample is $8.2$ points for
MolScribe, $9.7$ for plurality voting, $11.1$ for the union oracle, $11.6$ for
Sonnet~4.6 and $11.8$ for Opus~4.8. On the remaining $837$ molecules Opus~4.8 scores
$73.7\%$. We report full-sample numbers as the headline because they reflect the real
document distribution, but any comparison against work that excludes Markush structures
has to account for this.

\subsection{Relocated tables}
Tables~\ref{tab:taxonomy} and~\ref{tab:ablation}, referenced from
Sections~\ref{sec:where} and~\ref{sec:mechanism}, are placed here for space.

\begin{table}[t]
\caption{Failure taxonomy over the residual errors, mutually exclusive and assigned by
rule (Appendix~\ref{app:taxonomy}). Opus~4.8 removes constitutional errors and invalid
strings but produces more stereochemical ones.}
\label{tab:taxonomy}
\begin{center}
\small
\begin{tabular}{lccc}
\multicolumn{1}{c}{\bf Error class} & \multicolumn{1}{c}{\bf Sonnet 4.6}
& \multicolumn{1}{c}{\bf Opus 4.8} & \multicolumn{1}{c}{\bf Change}
\\ \hline \\
Markush R-group & 164 & 163 & -1 \\
Skeleton & 129 & 102 & -27 \\
Stereochemistry & 84 & 100 & +16 \\
Salt / counter-ion & 20 & 15 & -5 \\
No valid output & 6 & 1 & -5 \\
\hline \\
Total errors & 403 & 381 & -22 \\
\end{tabular}
\end{center}
\end{table}

\begin{table}[t]
\caption{Leave-one-reader-out, re-running the full pipeline on a fixed subset of
$n=300$ molecules. Under Sonnet~4.6 two of the three readers are load-bearing; under
Opus~4.8 no single reader is, and every deficit is within noise.}
\label{tab:ablation}
\begin{center}
\small
\begin{tabular}{llccccc}
\multicolumn{1}{c}{\bf Arbitrator} & \multicolumn{1}{c}{\bf Reader withheld}
& \multicolumn{1}{c}{\bf Alone} & \multicolumn{1}{c}{\bf Accuracy}
& \multicolumn{1}{c}{\bf $\Delta$ (pp)} & \multicolumn{1}{c}{\bf 95\% CI}
& \multicolumn{1}{c}{\bf $p$}
\\ \hline \\
Sonnet 4.6 & none (full) & --- & 62.0\% & --- & --- & --- \\
 & DECIMER & 44.7\% & 56.3\% & -5.7 & [-9.0, -2.7] & $<$0.001 \\
 & MolScribe & 45.0\% & 57.7\% & -4.3 & [-7.7, -1.3] & 0.011 \\
 & MolSight & 43.0\% & 60.3\% & -1.7 & [-5.0, +1.7] & 0.442 \\
Opus 4.8 & none (full) & --- & 63.0\% & --- & --- & --- \\
 & DECIMER & 44.7\% & 62.0\% & -1.0 & [-4.0, +2.0] & 0.664 \\
 & MolScribe & 45.0\% & 61.7\% & -1.3 & [-5.0, +2.3] & 0.585 \\
 & MolSight & 43.0\% & 62.3\% & -0.7 & [-4.0, +2.7] & 0.845 \\
\end{tabular}
\end{center}
\end{table}

\begin{table}[t]
\caption{Arbitration behaviour per 1000 molecules. Opus~4.8 opens a correction on
3.5$\times$ as many molecules, yet changes the seeded answer only slightly more often:
most of its extra proposals are talked down by its own validation gates.}
\label{tab:behaviour}
\begin{center}
\small
\begin{tabular}{lcc}
\multicolumn{1}{c}{\bf Quantity} & \multicolumn{1}{c}{\bf Sonnet 4.6}
& \multicolumn{1}{c}{\bf Opus 4.8}
\\ \hline \\
Molecules where any agent proposed a fix & 120 & 423 \\
Final answer differs from the seeded one & 253 & 278 \\
Correct although no reader was & 46 & 76 \\
Lost although some reader was correct & 57 & 65 \\
Knowledge-base entries written (total) & 13129 & 13648 \\
Knowledge-base entries per molecule (median) & 11 & 10 \\
\end{tabular}
\end{center}
\end{table}

\subsection{Integrity checks}
\label{app:integrity}

Before any analysis we verified: (i) the two conditions share all $1000$ molecule
indices; (ii) canonicalised base predictions are identical across the two conditions for
every molecule and every reader ($0$ mismatches), which is what makes the comparison
paired; (iii) recomputing each system's correctness from its stored final SMILES
reproduces the pipeline's own flag for every molecule in both conditions ($0$
disagreements); and (iv) exactly one ground-truth string in the sample fails to parse
under RDKit, and is counted wrong for all systems.

\subsection{Agent activity}

Under Sonnet~4.6 the knowledge base accumulates $13{,}129$ entries over $1000$ molecules
(median $11$ per molecule, maximum $52$); under Opus~4.8, $13{,}648$ entries (median
$10$, maximum $53$). Molecules the system ultimately gets wrong carry more discussion
than ones it gets right ($14.7$ versus $12.1$ entries under Sonnet~4.6), so
knowledge-base volume is a usable signal of difficulty. The StructuralAnalyst and the
StereoSpecialist write most entries, since they run on every molecule; the Chemist is
invoked on $101$ molecules under Sonnet~4.6 and $158$ under Opus~4.8. It is worth being
precise about where the $3.5\times$ intervention gap of Section~\ref{sec:mechanism} comes
from: counting by author, molecules carrying a proposed fix rise from $78$ to $143$ for the
Chemist but from $78$ to $377$ for the StructuralAnalyst. Under the stronger arbitrator it
is therefore mainly the agent that reads the image, rather than the agent designated to
repair the structure, that raises objections.

\end{document}